\documentclass[%
	final,
	notitlepage,
	narroweqnarray,
	inline,
	twoside,
	]{ieee}

\usepackage{amssymb}
\usepackage{graphicx}
\usepackage{url}

\newcommand{\noteperso}[1]{\begin{center}
 \fbox{\begin{minipage}{6cm}#1\end{minipage}}\end{center}}
\renewcommand{\noteperso}[1]{}

\begin{document}

\title[Measurement of eDonkey Activity
with Distributed Honeypots]{%
Measurement of {\em eDonkey} Activity\\
with Distributed Honeypots}

\author{
\begin{minipage}{12.3cm}
\center
Oussama Allali, Matthieu Latapy and Cl{\'e}mence Magnien\\
\medskip
\mbox{LIP6 -- CNRS and Universit{\'e} Pierre et Marie Curie (UPMC -- Paris 6)}\\
104, avenue du Pr\'esident Kennedy\\
75016 Paris -- France\\
\url{firstname.lastname@lip6.fr}
\end{minipage}
}

\maketitle               

\begin{abstract} 
Collecting information about user activity in peer-to-peer systems is a key but challenging task. We describe here a distributed platform for doing so on the {\em eDonkey} network, relying on a group of {\em honeypot} peers which claim to have certain files and log queries they receive for these files. We then conduct some measurements with typical scenarios and use the obtained data to analyze the impact of key parameters like measurement duration, number of honeypots involved, and number of advertised files. This illustrates both the possible uses of our measurement system, and the kind of data one may collect using it.
\end{abstract}


\noteperso{A FAIRE apres soumission:
\begin{itemize}
\item un papier sur la popularite des fichiers et/ou le blacklisting, pour TSI ?
\item mettre les donnees en ligne, anonymisees et en XML
\end{itemize}
}

\noteperso{A FAIRE avant soumission:
\begin{itemize}
\item uniformiser les noms de messages (y compris dans les figures)
\item dans les figures 2 et 3, mettre comme legendes des axes verticaux : "Total number of peers" et "Number of new peers"
\item dans les figures 5 et 6, ajouter comme legende de l'axe vertical : "Number of peers"; figure 7 ajouter comme legende verticale "Number of REQUEST-PART queries"
\item figure 10: honeypotS
\end{itemize}
}

\noteperso{ML: pour toutes les courbes ou on observe qqe chose de lineaire voir la pente de la droite dans le .tex original}

\section{Introduction.}

Peer-to-peer now accounts for the use of 70\% or more of the global
internet bandwidth~\cite{cisco,adsl}. It is therefore important to
understand the underlying mechanisms, in order to design efficient
protocols. In particular, user behaviors and data properties are of
high interest: are there correlations among the files owned by each
user? are there communities of interest? are there sets of similar
data or very popular data? how does data popularity evolve? ...

Much work has been done to answer these questions \cite{kasa,pam,isp}. The most widely used approach is to watch the activity in a network of interest for a period of time. This raises important difficulties due mainly to the distributed and anonymous nature of peer-to-peer systems, their dynamics, and their sheer size.

This paper explores a new method for collecting information on what
occurs in one of the main peer-to-peer systems currently in use, {\em
  eDonkey}\,\footnote{{\em eDonkey} is a semi-distributed peer-to-peer file
  exchange system based on directory servers. An unofficial
  documentation of the protocol is available~\cite{edonkey}.}.
It consists in introducing {\em honeypots} in the network,
 {\em i.e} peers pretending to offer files and logging
all the queries they receive for these files from other peers.

We begin by presenting the current state of the art of peer-to-peer measurement (Section~\ref{sec-related}), and then we describe
the new solution we propose (Section~\ref{sec-honeypots}). Finally we will illustrate its relevance
with some practical measurements (Section~\ref{sec-measurements}),
before presenting our conclusions (Section~\ref{sec-conclu}).

\section{Related work.}
\label{sec-related}
Measurement of peer-to-peer systems is a very active area of research. We focus here on the measurement of peer activity, thus ignoring measurements of peer-to-peer overlays and protocols, which are out of the scope of this paper.
Several approaches have been used to collect information on peer-to-peer activity, each with its own advantages and drawbacks. We rapidly describe them in this section.

\subsection{Measurement at server level.}
In centralized or semi-centralized systems (like {\em eDonkey}), queries
sent by peers are managed by servers; it is then possible to
passively collect data directly on these servers.
The measurement setting then consists
in either modifying the server so it may log the queries it manages
\cite{Glatapy,Clust}, or capturing traffic at IP level and then
decoding it \cite{hotp2p}.

This method has the advantage of collecting {\em all} the information
managed by a given server.
However, as actual file exchanges occur between peers (out of the sight of
servers), this information is not captured.
Moreover, it requires cooperation with server administrators.

\subsection{Measurement at peer level.}
Passive measurements are also possible in fully distributed systems at client level: 
a modified client may observe the traffic going through it including in some cases 
keyword queries, file searches, etc.

In~\cite{kasa,pam,Gnutone,isdev} authors set up such measurements. The
main issue of this approach is the need for users that agree to
cooperate, which limits the amount of data obtained. To
increase it, the author of~\cite{GnuDC}
designed a large distributed architecture called GnuDC (Gnutella
Distributed Crawler); it monitors the network by being attached to a
large number of peers.

\subsection{Measurement by client sending queries.}
An active measurement method from clients is also possible. It consists in designing a client that sends
queries in the system and records the obtained answers (lists of
files and providers, typically). This has been done in {\em Napster}~\cite{usrq}
and {\em eDonkey}~\cite{shre,clust2} with success.
The main drawback of this approach is that it is active: it may
interfere with the observations, and the rate at which queries may be
sent is limited.

\subsection{Measurement at ISP level.}
Finally, one may capture peer-to-peer traffic directly on ISP
infrastructures, in a passive way. 
  In
\cite{isp,tel,tom} for instance, data is collected from several routers,
and different peer-to-peer applications (Gnutella, FastTrack, DirectConnect,
{\em eDonkey}) are observed.

This approach provides network-level data, with limited information on users and exchanged files. This makes it quite different from other approaches discussed here.    
Moreover, it relies on cooperating
with ISPs, which have limited rights to observe user traffic.
\section{Our data collection system.}
\label{sec-honeypots}
Our measurement infrastructure consists in a
set of fake peers (the honeypots) connected to different servers, and
a manager controlling these honeypots. We first present our manager and honeypots,
 and then we discuss privacy concerns.
\subsection{Manager.}
The manager's role is to set up the honeypots,
and then coordinate them and centralize the data they collect.

The first function of the manager is to launch the honeypots. It
specifies to each of them a server to connect to.
Each honeypot then attempts to connect to the server, and reports its status (connected or not), as well as
its {\em clientID}\,\footnote{In the {\em eDonkey} network, peers are
  identified by a { \em clientID}, which is their IP address if they
  are directly reachable (high ID) or a 24 bits number otherwise (low ID). },
if relevant,
to the manager.
This makes it possible to re-launch dead honeypots or to redirect them
toward other servers. The manager regularly
checks the status of each honeypot for the same reason.

Several strategies make sense for assigning honeypots to servers.
One may
typically choose a different sever for each honeypot, in order to obtain
a more global view.
The choice of servers may also be guided by their
resources and number of users, so that the honeypots may reach the
largest possible number of peers.

The second function of the manager is to tell honeypots to
advertise fake files. It specifies the name, size and {\em fileID}\,\footnote{In the {\em eDonkey} network,
  files are identified by their {\em fileID} or {\em hash}. 
  It is generated from the file's content, which ensures that two
  files with the same content are considered as identical, regardless of their names.} of each file.
Again, many strategies are possible to choose the files to advertise,
and the manager is in charge of implementing the chosen strategy.
For instance, it is possible to study the activity on a specific topic by choosing 
 the files accordingly.
One may then ask to the discovered peers their list of shared 
files and add these files to the the honeypot's list. 
One may also advertise random files.

Finally, the manager periodically gathers the data collected by honeypots
and does some basic data management
(for instance, it merges and unifies the
collected log files, see below). 

\subsection{Honeypots.} 
After receiving an
order from the manager to advertise a file $F$ ,
the honeypot adds this file to its
shared file list.
The {\em eDonkey} server then adds the honeypot to its list  of 
providers for file $F$. 
A honeypot may therefore afterward be contacted by other peers  looking for $F$.
Notice that, due to peer exchange~\cite{kad} and other high-level {\em eDonkey} features, 
the honeypot may be contacted by peers which are not connected to the server.

Each honeypot constructs a log file of all the queries it receives.
The log file can be written directly on a hard disk or sent via
network to the monitor.

Before going any further, we specify that,
for each query, the honeypot saves the information contained in the {\em eDonkey}
protocol concerning the message type, as well as
meta-data such as the IP address, port, name, {\em userID}\,\footnote{The user ID (also called user hash) is unique and is used to identify a client across sessions,however,the client ID is valid only through a client's session with a specific server~\cite{edonkey}. }, 
version of client and ID status (high or low)
of the peers sending the queries;
moreover, it collects information concerning the server
(name, IP, port),
 as well as data concerning the
network environment (such as the timestamps marking the reception of the
packets by the system).

Once a peer is connected to a honeypot, the list and description of all
shared files in its cache content are retrieved. Note that this feature is not available
on all peers, as it can be disabled by the user.

\medskip
A honeypot must pass for a normal peer on the network. For this
purpose, we have modified the open-source {\em eDonkey} client
Amule~\cite{amule} so that it meets our
needs. 
We detail the main
modifications  below.
\begin{description}
\item[File display.]
In the normal course of events, if a client has files
to offer (the client application considers that all files belonging
to a given directory are files to be shared with other users),
an OFFER-FILES message describing these files is sent immediately after the
connection to the server is established,
or whenever the client's shared file list changes.
This message can also be used as a
{\em keep-alive} message, sent periodically to the server. 

As our honeypots do not have any file in their shared file folder, we have added a module to
change the shared file list,
so that the honeypot automatically sends the desired OFFER-FILES messages to the server.
\item[Communication with other peers.] When a honeypot is connected to a server and advertises some files,
it may be contacted by other peers wanting to download one of these files.
The normal message
exchange process between a peer wanting to download a file  and a peer having the file
is presented in Figure~\ref{fig-file-query}. 
The downloading peer tries to establish a connection
with a HELLO message,
to which the provider peer answers with a HELLO-ANSWER message.
The downloading peer sends a
START-UPLOAD query, declaring which file it is interested in;
 the provider replies with an
ACCEPT-UPLOAD message.
The downloading peer then starts to query file parts  and
the provider sends the queried parts.

\begin{figure}[h!]
  \centering
  \includegraphics[width=4.5cm]{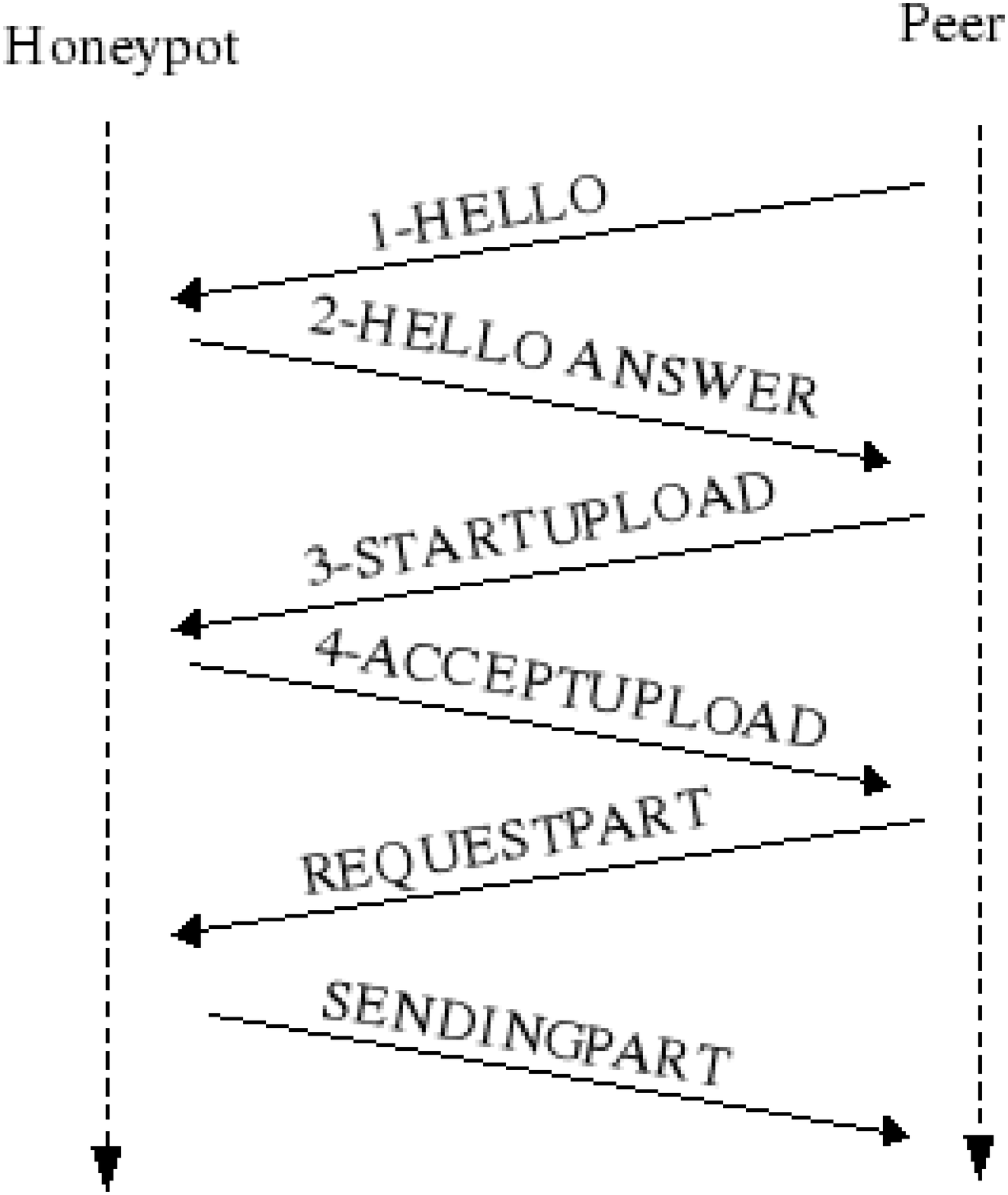}
  \caption{Series of messages exchanged between honeypots and peers.}
  \label{fig-file-query}
\end{figure}

Our honeypots behave like provider peers up to
the last part of the exchange: they do not send parts of the desired file
to the downloading peer.
In this aspect, honeypots do therefore not act as normal
peers in the system, and run the risk of being noticed and then blacklisted.
To avoid this,
we have implemented two different strategies for answering file parts 
queries.
The first one consists in sending nothing:
the corresponding honeypots do not reply to file parts queries.
The other strategy consists in sending random content when queried
for file parts.
\item[Log file construction.]Our goal is to record users' activity concerning the files a honeypot
advertises.
We have therefore modified the application to record the following
message types received from other peers: HELLO, START-UPLOAD, and REQUEST-PART. 
\end{description}
\subsection{Privacy concerns }
For ethical and legal reasons, we cannot record any 
sensitive data concerning users' privacy.
In our context, the main such data are IP addresses and filenames.
We want however to be able to know if a same user requested several
files or not.

In order to attain a high level of anonymisation, while keeping the anonymised
data coherent, we follow a two-steps procedure.
First, each
honeypot encodes IP addresses in its log using a one way hash function,
cryptographically sure. This anonymisation takes place before any data is written
to disk or sent to the manager.
This is however not sufficient to achieve a secure anonymisation of IP addresses:
somebody could apply the hash function to all $2^{32}$ possible IP addresses,
to construct a reverse dictionary of the anonymisation.
After having collected the data from the honeypots,
the manager therefore conducts a second anonymisation step:
it replaces each hash value, in a
coherent way between honeypots' logs,
by an integer: the first hash is replaced by 0, the second one by 1, and so on.
This ensures that the final anonymised data is secure,
and that it is not possible to obtain the users' IP addresses from it.

In addition, we also anonymise file names which may contain personal 
information~\cite{file-inf,Adar2007}. 
We replace each word that appears less often than a given threshold by an integer.

\section{Experiments.}
\label{sec-measurements}

Many parameters may have a strong impact on the measurements conducted with our tool: which files the honeypot claims to have; number of honeypots in a distributed measurement; number of files advertised by the honeypot; duration of the measurement; etc.

In this section, we study several measurements which we have conducted to illustrate what is feasible with our approach and tool, and to investigate the impact of some key parameters. This relies on two different measurements:
\begin{itemize}
\item
The {\em distributed} measurement used 24 honeypots ran by different PlanetLab~\cite{plant} machines during one month (October 2008). Among the 24 honeypots, half did not answer at all to queries received from other peers; the others sent files with random content to the peers contacting them (see Section~\ref{sec-random} below). All honeypots advertised the same four files (a movie, a song, a linux distribution and a text). They were all connected to the same large server and had a {\em HighID}.
\item
The {\em greedy} measurement used only one honeypot but aimed at advertising as many files as possible. To do so, the honeypot ran a two-steps procedure: during the first day of measurement, it asked their list of shared files to all peers contacting it, and added all these files to its own list of shared files; after the first day, it did not increase its list of shared files anymore, and just recorded the queries it received and the lists of files shared by peers contacting it. It did not send any content to other peers. We ran this measurement during the two first weeks of November 2008.
\end{itemize}
The key properties of the obtained data are summarized in Table~\ref{tab-basic}. These statistics already show that our measurement method succeeds in collecting large amounts of data, with hundreds of thousands distinct peers and files observed.

\begin{table}[h!]
\centering
\begin{tabular}{|c|c|c|}
\cline{2-3}
\multicolumn{1}{c|}{} & {\bf distributed} & {\bf greedy}\\
\hline
Number of honeypots & 24 & 1\\
\hline
Duration in days & 32 & 15\\
\hline
Number of shared files & 4 & 3,175\\
\hline
Number of distinct peers & 110,049 & 871,445\\
\hline
Number of distinct files & 28,007 & 267,047\\
\hline
Space used by distinct files & 9 TB & 90 TB\\
\hline
\end{tabular}
\caption{Basic statistics regarding the data collected with our measurements.}
\label{tab-basic}
\end{table}

In the following, we investigate the impact of four key parameters of
the measurement: its duration; the fact that honeypots send random
content or no content at all; the number of honeypots involved; and
the number of advertised files.

\subsection{Impact of measurement duration.}
\label{sec-duration}

\begin{figure}[h!]
\centering
\includegraphics[width=8.45cm]{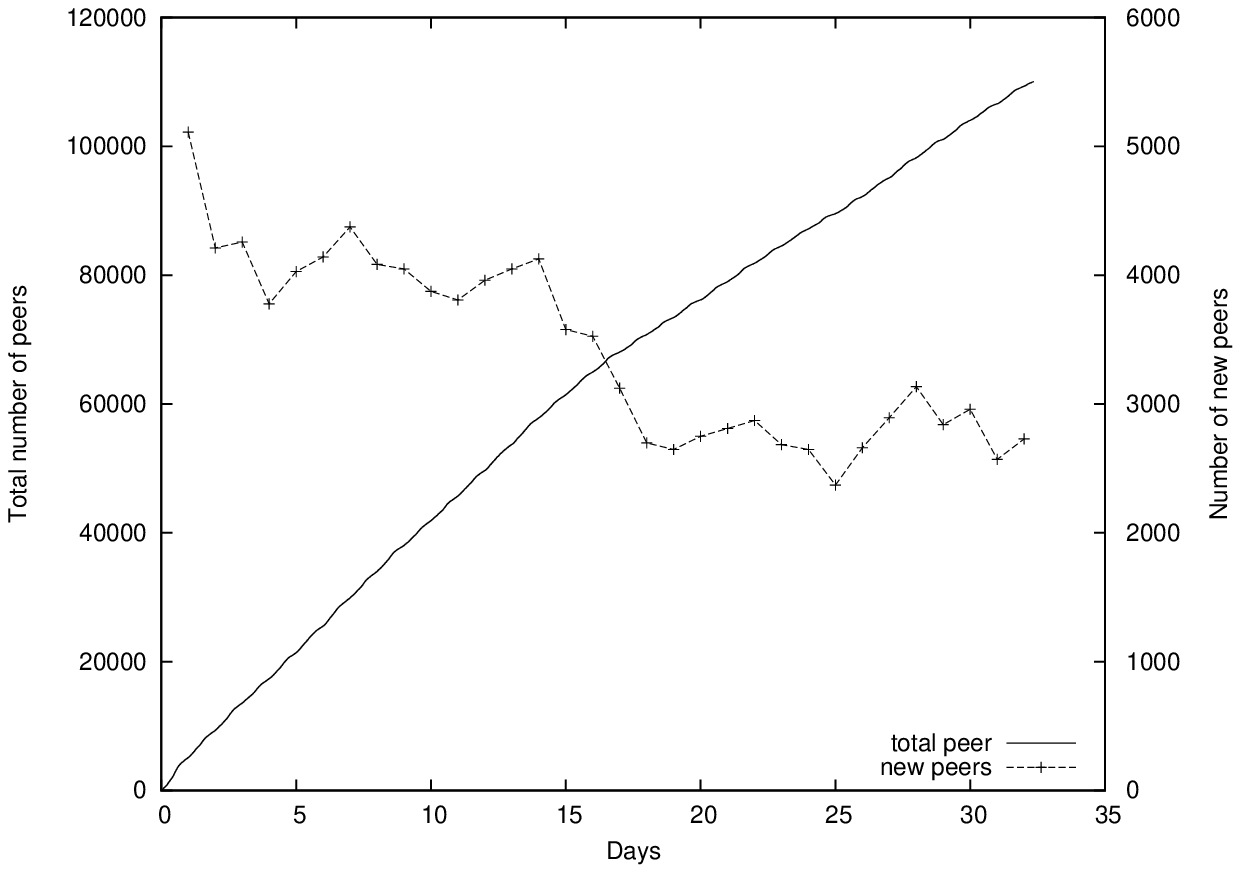}
\caption{Evolution of the number of distinct peers observed during our distributed measurement (left vertical axis) and number of new peers observed each day (right vertical axis) as a function of time elapsed since the beginning of this measurement, in days (horizontal axis).}
\label{fig-nbpeers-distributed}
\end{figure}

\begin{figure}[h!]
\centering
\includegraphics[width=8.45cm]{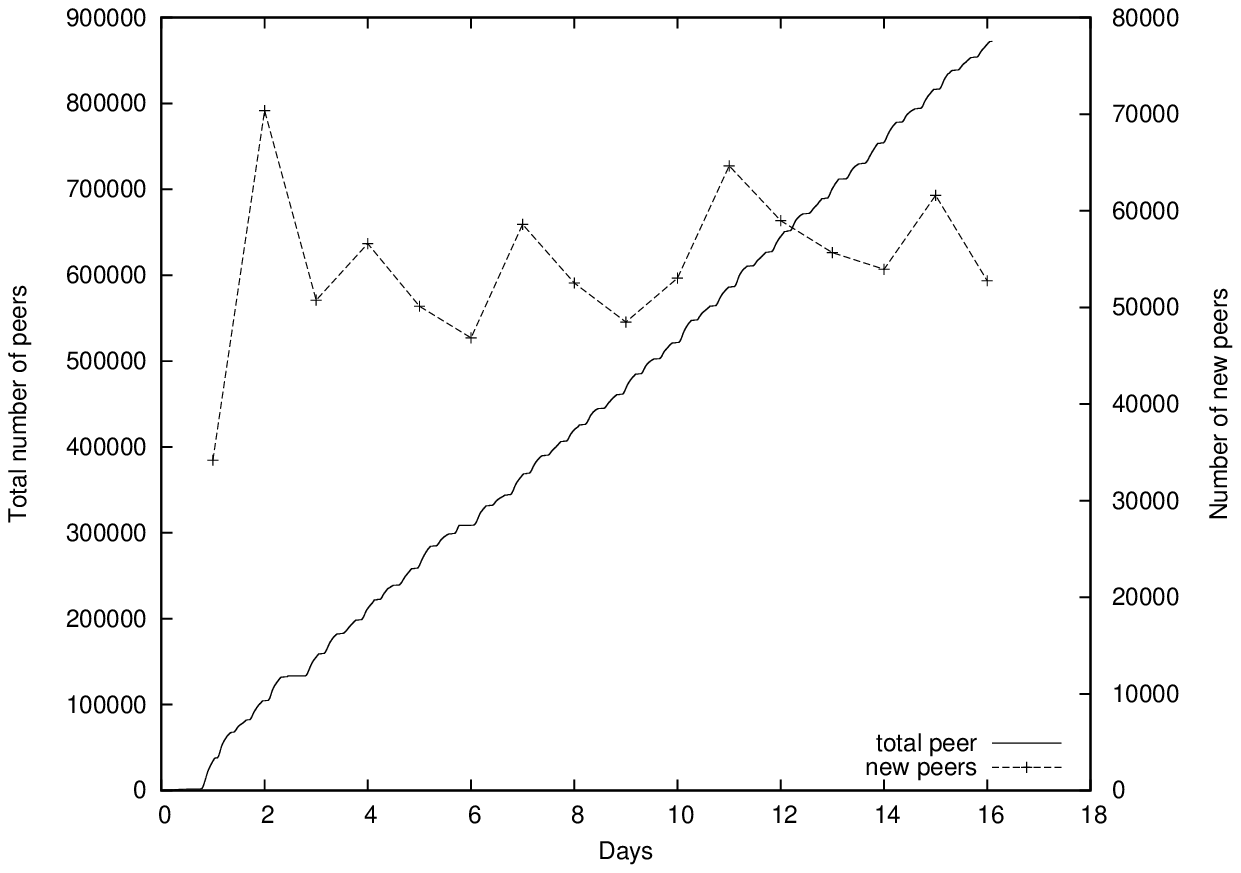}
\caption{Evolution of the number of distinct peers observed during our greedy measurement (left vertical axis) and number of new peers observed each day (right vertical axis) as a function of time elapsed since the beginning of this measurement, in days (horizontal axis).}
\label{fig-nbpeers-greedy}
\end{figure}

Let us first observe how the number of distinct observed peers evolves as the duration of the measurement grows, displayed in Figures~\ref{fig-nbpeers-distributed} and~\ref{fig-nbpeers-greedy} for the distributed and greedy measurements respectively. It appears clearly that the number of observed peers grows rapidly, and linearily, during all the measurement. Even after 30 (resp. 15) days, the number of distinct peers observed by the distributed (resp. greedy) measurement still grows significantly: more than $2,500$ (resp. $50,000$) new peers per day.

These observations are of prime importance for conducting measurements: they show that conducting very long measurements is relevant, as one continuously discovers a significant amount of new peers. They also show that blacklisting, even if it is present, does not prevent us from observing many peers despite the fact that our honeypots never provide any useful content. We deepen this in the next section.

The number of new peers discovered each day is also displayed in Figures~\ref{fig-nbpeers-distributed} and~\ref{fig-nbpeers-greedy}. It decreases during time in the distributed measurement, which is probably due to the fact that the popularity of shared files decreases. This may also be due to the fact that we reach a situation where most peers interested in the files proposed by the honeypot already have contacted it. In any case, the important point here is that this happens only after a very long measurement duration (one month), and that even then the number of new observed peers remains large (more than $2,500$ per day). This shows that continuing the measurement for long periods of time makes sense, even with $24$ distributed honeypots advertising only $4$ files. In such a scenario, one may have guessed that all the peers potentially interested in the proposed files, or most of them, would have been observed before $30$ days. Of course, this depends on the popularity of the proposed files, but deepening this is out of the scope of this paper.

One may also notice that the number of peers observed during the first day of our greedy measurement (Figure~\ref{fig-nbpeers-greedy}) is very low; this is due to the initialisation phase of this measurement strategy, as described above: during the first day, the honeypot mainly constructs its large list of shared files, starting with only a few. The number of observed peers during this period of time is non-zero, but it is much smaller than after the initialisation phase and thus it is not visible on the plot. After this initial period, the honeypot observes an average of $54,000$ new peers each day, which is stable during the $15$ days of measurement.

\begin{figure}[h!]
\centering
\includegraphics[width=8.45cm]{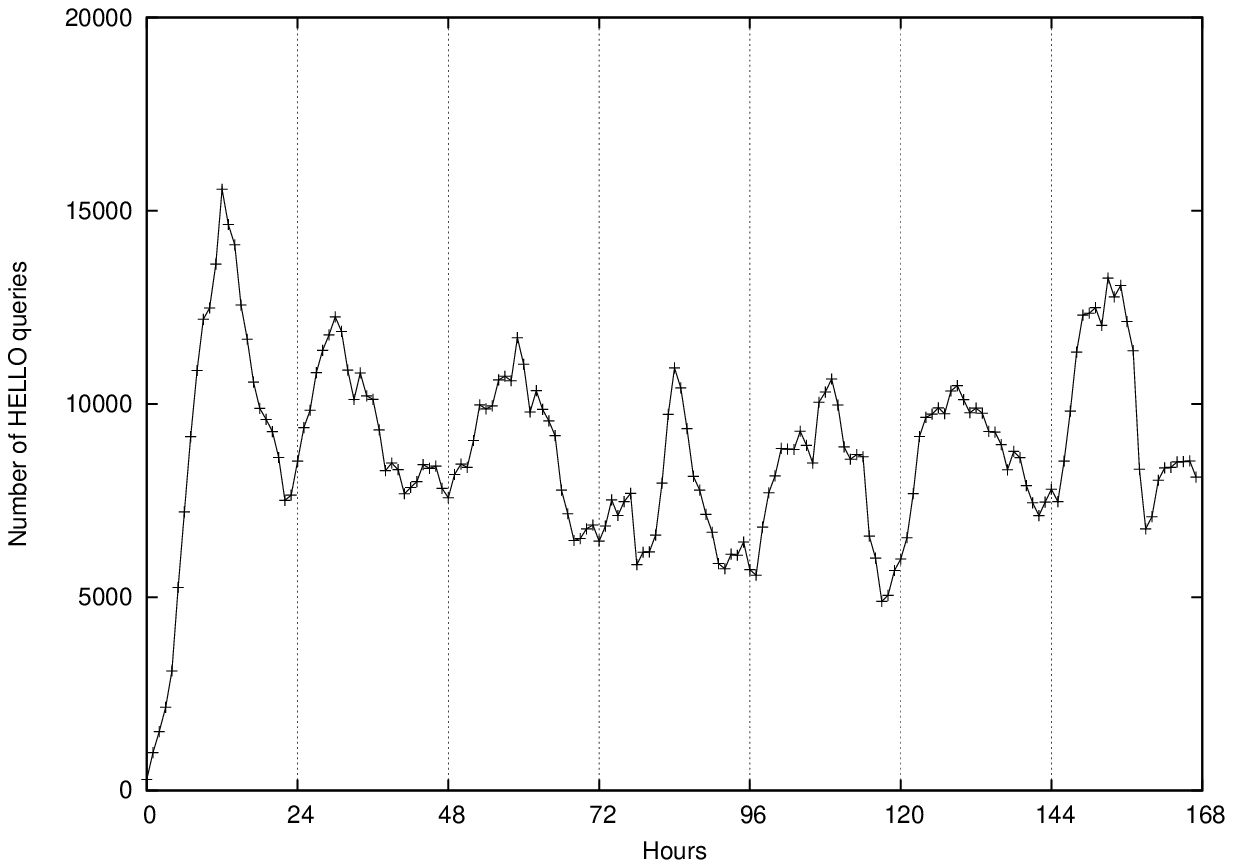}
\caption{Number of HELLO messages received each hour   during the first
  week of our distributed measurements (vertical axis)
  as a function of  time elapsed
  since the beginning of the measurement, in hours (horizontal axis).}
\label{fig-nbpeers-time}
\end{figure}

Finally, we display in Figure~\ref{fig-nbpeers-time} the number of
queries we received during each hour of the first week of our
distributed measurement (which is representative of other weeks and
other measurements). After an initial phase in which our honeypots are
not well known in the system (it lasts 10 minutes before we get our
first query), a clear day-night effect appears. Such an effect
reflects a regional nature of {\em eDonkey} activity: if observed users were
scattered worldwide, such a day-night phenomenon would not appear. As
these day-night variations follow the European and North African daily
life variations (maximal values are reached during daytime in these
regions and minimal ones during night), the plot may be seen as a
confirmation of the well known fact that the {\em eDonkey} system is most
popular in these regions \cite{fed1,fed2}. This may also be a
consequence of a regional interest for the files proposed by the
honeypot (for instance, music and movies are subject to such effects).

\subsection{Random content vs no content.}
\label{sec-random}

When contacted by peers wanting an  advertised file, honeypots may
apply two different strategies: they may simply ignore these queries
and not answer them; or they may send random content. These strategies may play an
important role in avoiding blacklisting at server and/or peer levels:
if the system detects that honeypots do not provide relevant
content, then other peers may stop contacting them. We do
  not consider the strategy consisting in providing the true files,
  which would raise bandwidth and storage problems, as well as legal
  and ethical issues in many cases.

\begin{figure}[h!]
  \centering
  \includegraphics[width=8.45cm]{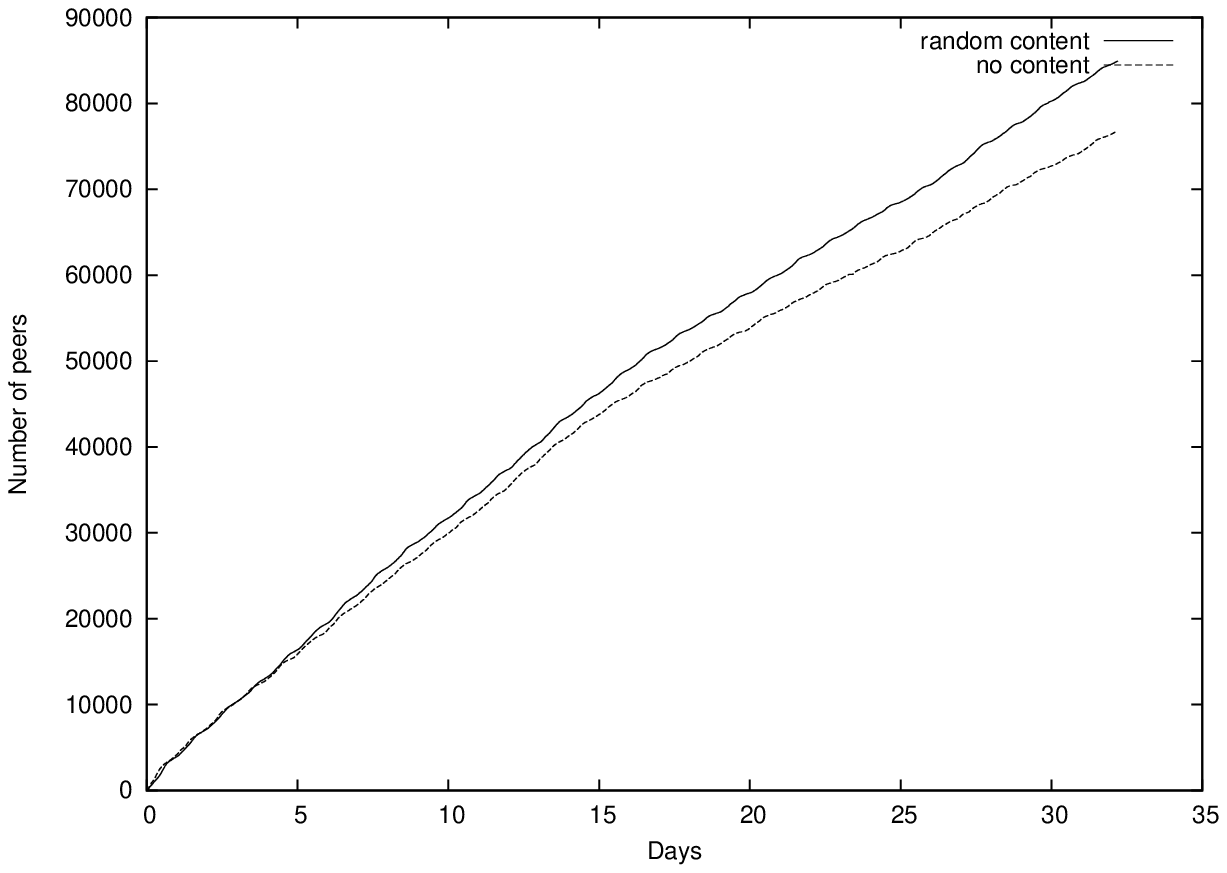}
  \caption{Number of distinct peers sending HELLO messages to our two
    groups of honeypots in our distributed measurements (vertical axis) as a function of time elapsed
    since the beginning of measurement, in days (horizontal axis).}
  \label{fig-hello}
\end{figure}

\begin{figure}[h!]
  \centering
  \includegraphics[width=8.45cm]{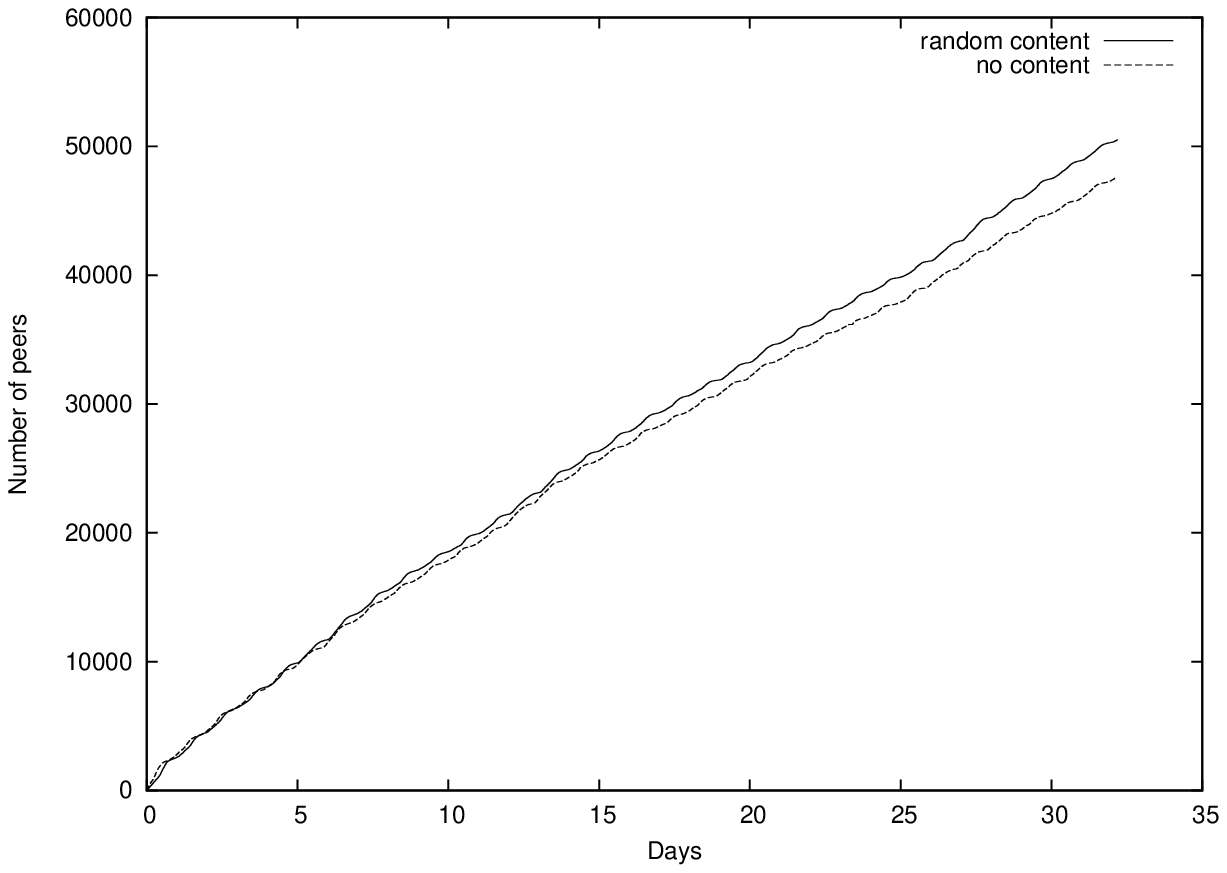}
  \caption{Number of distinct peers sending START-UPLOAD messages to
    our two groups of honeypots in our distributed measurements (vertical axis) as a function of time
    elapsed since the beginning of measurement, in days (horizontal
    axis).}
  \label{fig-upload}
\end{figure}

\begin{figure}[h!]
  \centering
  \includegraphics[width=8.45cm]{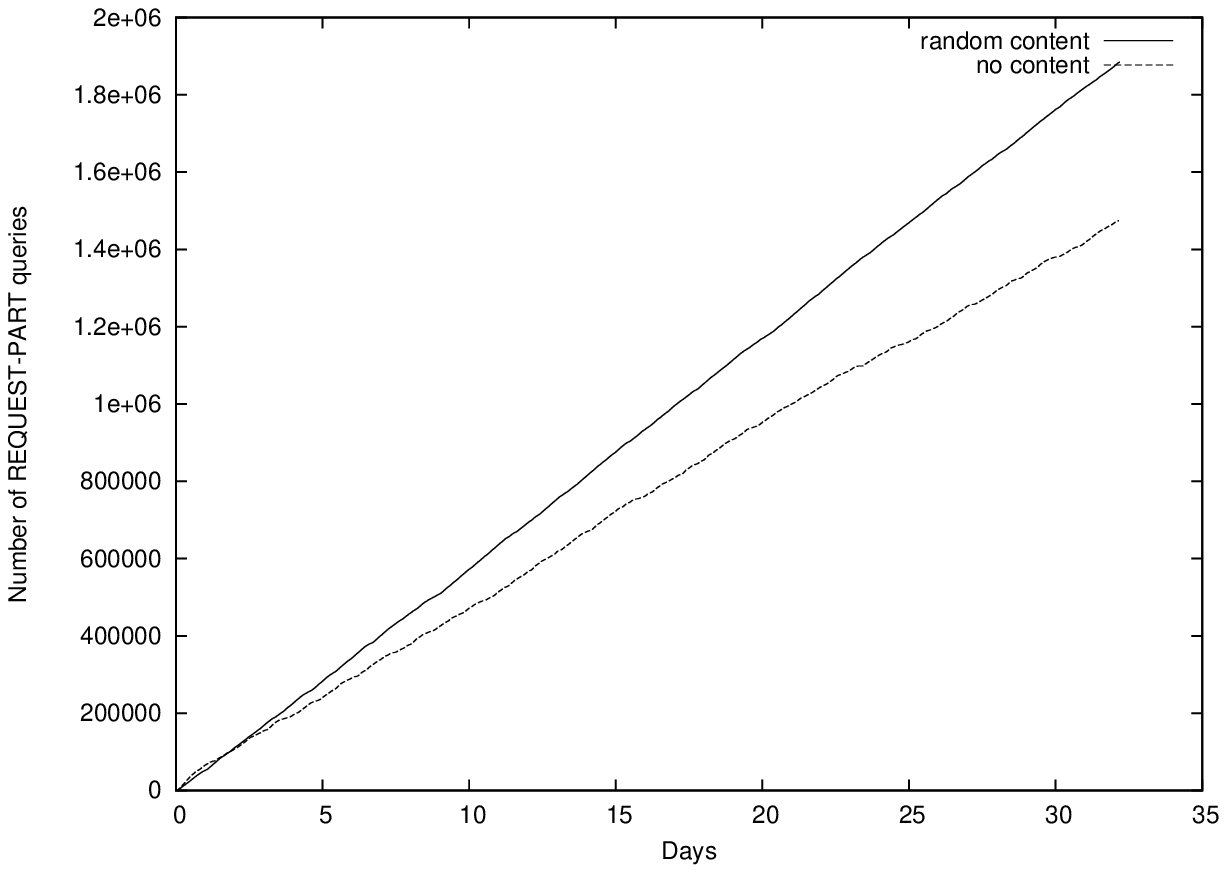}
  \caption{Number of REQUEST-PART messages received by our two groups
    of honeypots in our distributed measurements (vertical axis) as a function of time elapsed since
    the beginning of measurement, in days (horizontal axis).}
  \label{fig-part}
\end{figure}

In order to investigate this, half the honeypots in our distributed measurement applied the first strategy, and half applied the second one. This leads to two groups of $12$ honeypots, which we call {\em no-content} and {\em random-content}, respectively. Figures~\ref{fig-hello} and~\ref{fig-upload} show the number of distinct peers sending HELLO and START-UPLOAD messages (see Figure~\ref{fig-file-query}) observed by each group during our measurement. Similarily, Figure~\ref{fig-part} displays the number of REQUEST-PART messages they received. Similar plots are obtained for each file advertised by our honeypots, independently (recall that the honeypots of this measurement setup advertised 4 different files).

All these plots clearly show that, although the difference is not
huge, the {\em random-content} strategy leads to better results than
the {\em no-content} one. This is particularly striking regarding the
HELLO and START-UPLOAD messages, as the two strategies behave exactly
in the the same up till this point. This demonstrates that, even though we
still discover many new peers, there is some kind of blacklisting of
honeypots, and that this blacklisting is more efficient when the
honeypot sends no content. This is probably due to the fact that
detecting honeypots which send invalid content takes more time than
detecting honeypots which send nothing.

The larger difference between strategies regarding the number of
REQUEST-PART messages, as observed in Figure~\ref{fig-part} (we
finally observe $1,9$ million queries with the random-content
strategy, and only $1,5$ million with the no-content strategy), may be
explained as follows. First, when a peer receives irrelevant data or
no data from our honeypots, it may stop using it.
Detecting the fact that the honeypot sends random content is also longer
than detecting that it does not answer, and thus a peer may send
queries for more file parts in this case before deciding not to
consider the honeypot anymore. This may be seen as an implicit
blacklisting at client level, which stops using a honeypot after it
observed that it sends no useful content.

Finally, these plots show that the {\em random-content} strategy
is significantly more efficient than the {\em no-content} one. Other
blacklisting techniques may operate, but we cannot observe them with
our data.

\medskip

\begin{figure}[h!]
  \centering
  \includegraphics[width=8.45cm]{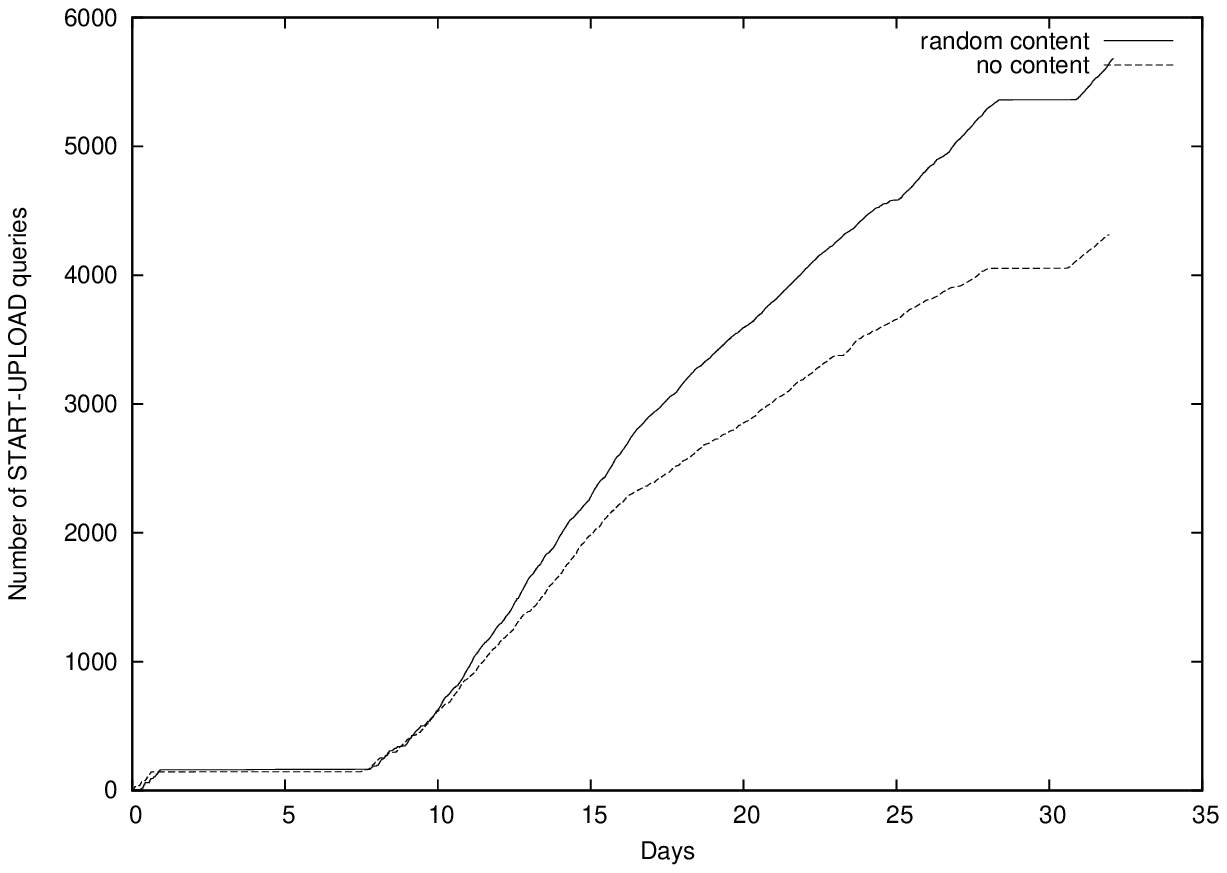}
  \caption{Number of START-UPLOAD messages received {\em from a single
      peer} by our two groups of honeypots  in our distributed measurements (vertical axis) as a
    function of time elapsed since the beginning of measurement, in
    days (horizontal axis).}
  \label{fig-start-one}
\end{figure}

\begin{figure}[h!]
  \centering
  \includegraphics[width=8.45cm]{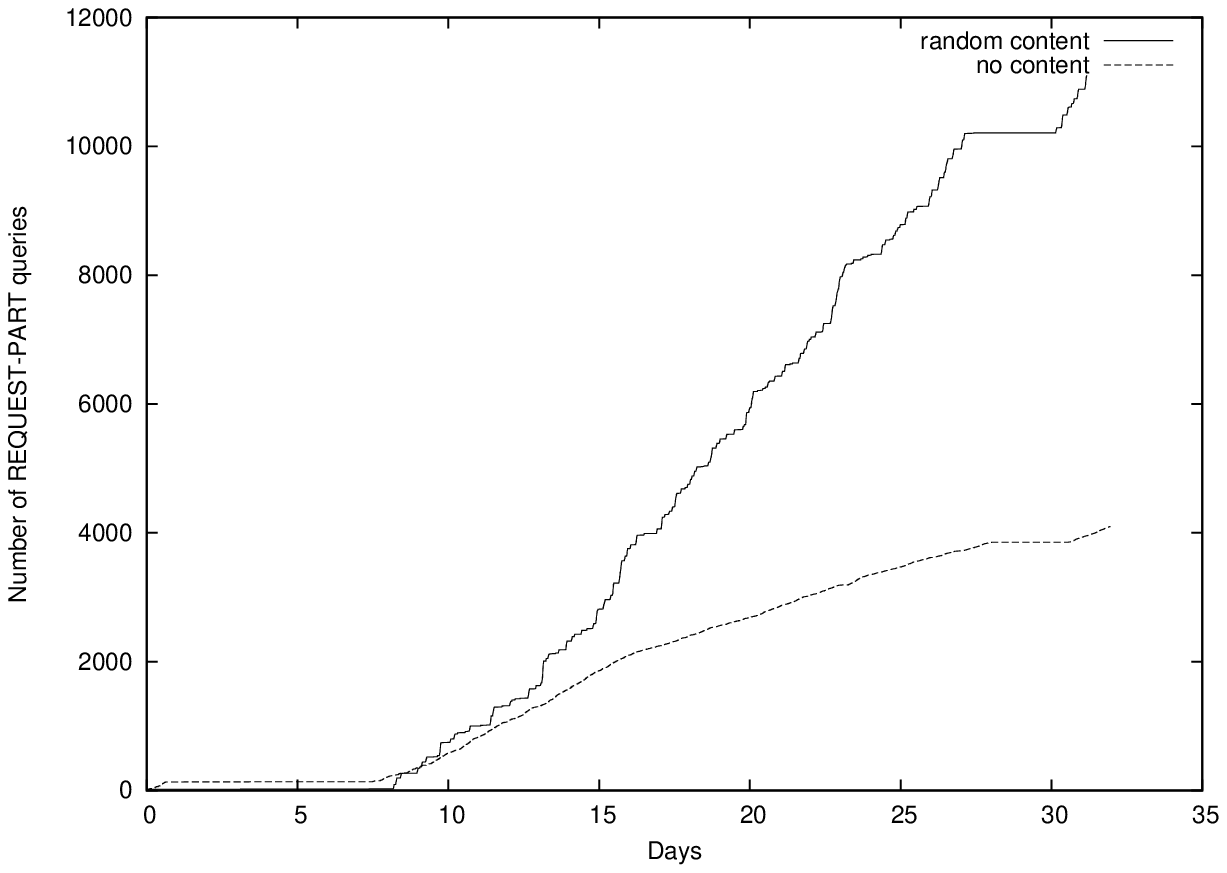}
  \caption{Number of REQUEST-PART messages received {\em from a single
      peer} by our two groups of honeypots  in our distributed measurements (vertical axis) as a
    function of time elapsed since the beginning of measurement, in
    days (horizontal axis).}
  \label{fig-part-one}
\end{figure}

In order to investigate further the difference between the {\em
  random-content} and {\em no-content} strategies, we plot in
Figure~\ref{fig-start-one} (resp. Figure~\ref{fig-part-one}) the
number of START-UPLOAD (resp. REQUEST-PART) messages received by our
two groups of honeypots from a {\em single} peer. The plot for HELLO
messages is very similar to Figure~\ref{fig-start-one} so we do not
reproduce it here.

The peer we have chosen is the one which sent the largest number of queries to our honeypots. For some periods of time, it does not send any queries (which induces a plateau on the plots), but in general this peer sends queries as fast as it can, provided that the previous query was finished before it sends the next one.

The fact that it sends significantly less queries to our {\em no-content} honeypots than to our {\em random-content} ones confirms that the queries which receive no answer are sent at a lower rate than the ones which receive random content, as claimed above. This also explains the fact that the plot for {\em no-content} is smoother than the one for {\em random-content}, see Figure~\ref{fig-part-one}: the time between two queries to a {\em no-content} honeypot is constant (it is the timeout of the peer waiting for an answer), while for {\em random-content} this time may vary.

Notice however that this is not sufficient to explain the difference between the two strategies observed in Figures~\ref{fig-hello} and~\ref{fig-upload}, as they display the number of distinct peers (not messages) observed. As explained above, this is certainly due to a combination of some kind of client-level blacklisting combined to the difference of speed at which a query is processed when sent to {\em no-content} and {\em random-content} honeypots.

\subsection{Impact of the number of honeypots.}
\label{sec-nbhoneypots}

Our tools makes it possible to conduct honeypot measurements from a large number of machines at the same time, in a distributed manner. In this section, we explore the benefit of this feature: one may imagine that increasing the number of honeypots does not improve the measurement, at least as soon as a quite small number of honeypots are in place. The key question we want to address here therefore is: given a number $n$ of honeypots, what is the benefit of adding one more honeypot to the infrastructure.

\begin{figure}[h!]
\centering
\includegraphics[width=8.45cm]{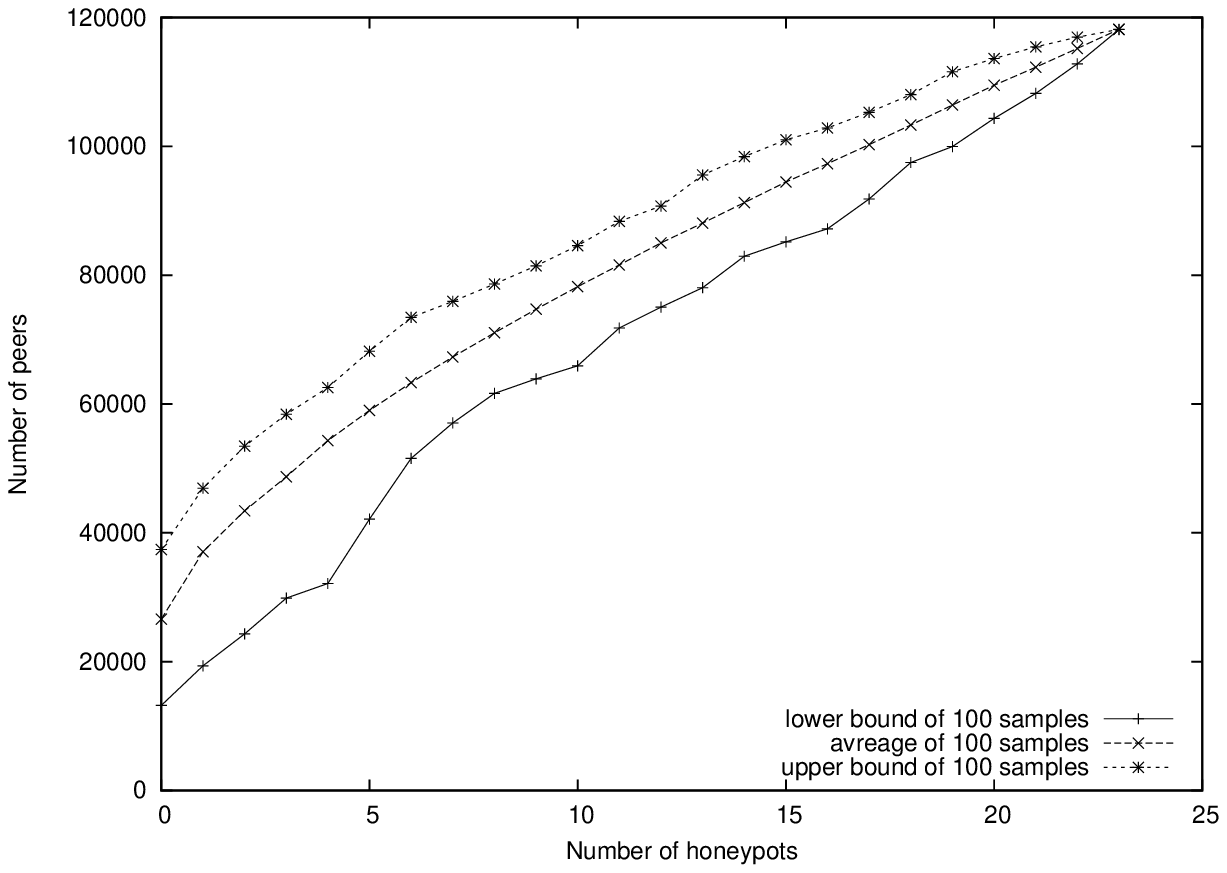}
\caption{Number of distinct peers observed at the end of the
  measurement (vertical axis) as a function of the number $n$ of
  involved honeypots (horizontal axis). For each $n$, we sample $100$
  random sets of $n$ honeypots and plot the average, minimal and
  maximal obtained values.}
\label{fig-nbhoneypots}
\end{figure}

To investigate this, we use our distributed measurement involving $24$ honeypots, and explore what we would have seen if only a part of them were used. To do so, we select $n$ random honeypots among the $24$ ones, for $n$ between $0$ and $24$, and we compute the number of distinct peers observed by these $n$ honeypots alone. As the choice of the $n$ honeypots may have a significant influence on the result, we repeat this computation $100$ times (we would ideally consider {\em all} $2^{24}$ possible subsets of our honeypots, but this is not feasible), and we plot the average, maximal, and minimal obtained values, see Figure~\ref{fig-nbhoneypots}. The average makes it possible to observe what one may expect. The maximal and minimal values give an indication of the dispersion of obtained results. In this case, they are quite close to the average, except at the very beginning of the plot (a honeypot leads to the observation of as many as $37,000$ peers, while another only sees $13,000$).

\noteperso{ML: figure~\ref{fig-nbhoneypots} : ca se termine par une droite ou pas ? moi j'ai dit que non dans le texte, mais a voir ; tracer un fit ? tracer le NB de nouveaux peers par honeypot ?}

This plot clearly shows that there is an important benefit in using a few honeypots rather than only one. It also shows that, even when $24$ honeypots are used, there is a significant benefit in adding more honeypots, thus calling for the use of large-scale distributed infrastructures. However, the benefit obtained by using more honeypots progressively decreases. This indicates that, even though many more honeypots may be used, at some reasonable point the benefit will become very low. Exploring this further requires distributed measurements from many more machines, which we leave for future work.

\subsection{Impact of the number of files.}
\label{sec-nbfiles}

A natural way to increase the number of peers observed by our honeypot
measurements is to increase the number of files we advertise, thus
reducing the focus of the measurement. This is the aim of our {\em
  greedy} measurement, which starts by collecting for one day the list
of files shared by the peers contacting a honeypot, and then adds all
these files to the list of files shared by the honeypot, as described
above.

Similarly to the previous section, we consider a measurement with a given number of files, and then, for any $n$ lower than or equal to this number, we study the number of peers we would have observed if we used only $n$ of these files in our measurement. Here, we could in principle consider the $3,175$ files advertised in our measurement. However, this is not feasible in practice; we therefore consider two subsets of files: the {\em random-files} are a set of $100$ randomly chosen files; and the {\em popular-files} are the $100$ files for which we received queries from the largest number of peers.

\begin{figure}[h!]
\centering
\includegraphics[width=8.45cm]{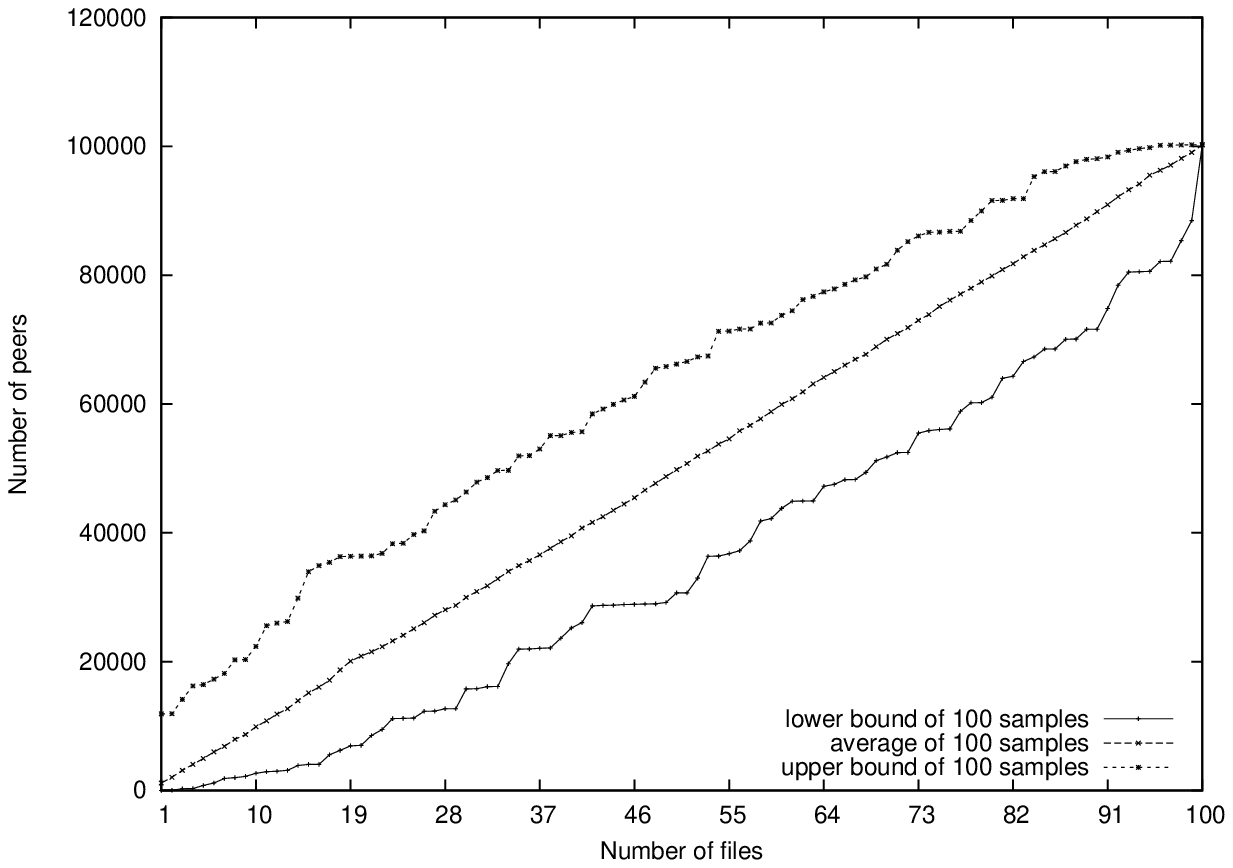}
\caption{Number of distinct peers observed at the end of the greedy measurement (vertical axis) as a function of the number $n$ of advertised files in the {\em random-files} set (horizontal axis). For each $n$, we sample $100$ random sets of $n$ files and plot the average, minimal and maximal obtained values. In average, each new file leads to the discovery of approximately $1,000$ new peers.}
\label{fig-file-rand}
\end{figure}

\begin{figure}[h!]
\centering
\includegraphics[width=8.45cm]{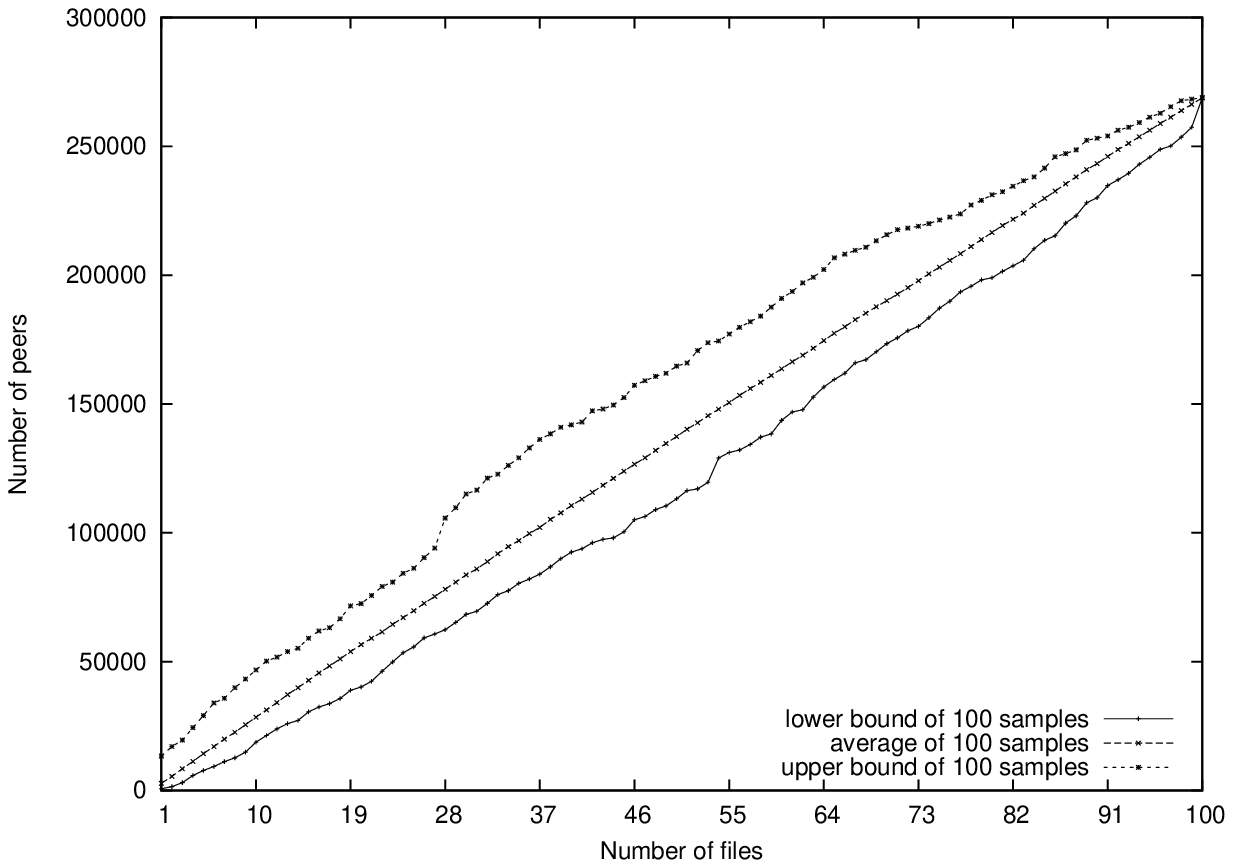}
\caption{Number of distinct peers observed at the end of the greedy measurement (vertical axis) as a function of the number $n$ of advertised files in the {\em popular-files} set (horizontal axis). For each $n$, we sample $100$ random sets of $n$ files and plot the average, minimal and maximal obtained values. In average, each new file leads to the discovery of approximately $2,700$ new peers.} 
\label{fig-file-pop}
\end{figure}

For these two sets of files, we select $n$ random ones among the $100$ in the set, for $n$ between $0$ and $100$, and we compute the number of distinct peers sending queries for at least one of these files. As the choice of the $n$ files may have a significant influence on the result, we repeat this computation $100$ times (we would ideally consider {\em all} $2^{100}$ possible subsets of files, or even the $2^{3\,175}$ ones, but this is not tractable), and we plot the average, maximal, and minimal obtained values, see Figures~\ref{fig-file-rand} and ~\ref{fig-file-pop}.

In both cases, the average, minimal and maximal values behave
similarly, but the difference in the number of peers observed at the
end of the measurement ($100,000$ for {\em random-files} and $270,000$ for
{\em popular-files}) clearly shows that the advertised files have a
significant impact on the obtained data. This is confirmed for any
number of files; in particular, the most popular file leads to the
discovery of $13,373$ peers, while some files lead to the observation
of $2$ peers only.

Importantly, these plots also show that the number of peers increases significantly and linearily as we add more files to our shared files list. This shows that the greedy approach certainly is relevant if one aims at observing huge number of peers, and that advertising more than $100$ files definitively is relevant.

\section{Conclusion.}
\label{sec-conclu}

We have described a method for peer-to-peer measurement via honeypots and implemented it on the {\em eDonkey} network. This method makes it possible to collect large amounts of data on peers searching for files. It complements other approaches and has its own advantages and drawbacks: it makes it possible to focus on a topic of interest, or even a specific file; it does not require cooperation with a server administrator and/or an ISP; but it gives a very partial view; and it may interfere with the systems as it is active.

We have illustrated the use of our approach by conducting some measurements which show that it is relevant: it captures information on many peers and files (hundreds of thousands) during long periods of time (several weeks), and using a relatively large number of distributed honeypots (24).

Using these measurements, we have evaluated several strategies and parameters. First, we studied the impact of measurement duration, and showed that it is useful to make very long measurements, as we continuously observe new peers and files. Then we compared the results obtained by sending random content to other peers {\em vs} not answering to their queries; there is no huge difference between the two, but sending random content is more efficient. We then studied the impact of the number of honeypots involved in a measurement, and the impact of the number of files they advertise; we showed that increasing both makes measurements much more efficient in terms of the number of observed peers and files.

Finally, the global conclusion is that it is relevant to conduct large-scale measurements in terms of both duration, number of involved honeypots, and number of advertised files.
Presented results may also be seen as giving hints in the setting up of measurements using our infrastructure, depending on one's priority and available resources.

The most immediate perspective for future work therefore is to conduct measurements at a significantly larger scale.
Another key direction is the design of a measurements targeting specific properties, like for instance being able to capture {\em all} the activity regarding a particular file or a set of files, and/or a specific keyword. The results presented here show that this is a difficult task: measurements advertising only 4 files and ran on 24 honeypots during one month do capture much data, but they do not reach a steady state.

More generally, designing strategies aimed at capturing different kinds of information is an important perspective: which files should honeypots advertise? how should distributed honeypots be coordinated? how many of them are necessary? ...

Analysis of the obtained data should also be deepened. In particular, we plan to explore the relationships between peers inferred from the fact that they are interested in the same files, and conversely study relations between files from the fact that they are downloaded by the same peers.

\smallskip
\noindent
{\bf Acknowledgements.}
This work is supported in part by the European MAPAP SIP-2006-PP-221003 and
French ANR MAPE projects.

\bibliographystyle{IEEEtran}
\bibliography{ieeecls}
\end{document}